# El físico Enrique Loedel Palumbo en el corredor científico Montevideo-Buenos Aires-La Plata: 1920-1930

**Alejandro Gangui**[1]
Universidad de Buenos Aires y CONICET

**Eduardo L. Ortiz**[2]
Imperial College London

*En este trabajo consideramos la atmósfera progresista y liberal del Montevideo de fines del siglo XIX y en ella la trayectoria del joven estudiante de ciencias Enrique Loedel Palumbo. Nos ocupamos luego de sus actividades en Argentina, donde se trasladó para estudiar Física en la Universidad de La Plata, en un nuevo y excelentemente equipado Instituto de Física en el que, eventualmente, llegaría a ser una figura muy destacada. Inicialmente Loedel Palumbo se ocupó del estudio de la estructura de moléculas complejas con base en sus propiedades ópticas, eléctricas y magnéticas. Más tarde tuvo la oportunidad de dialogar con Einstein cuando este visitó Argentina en 1925. Como resultado de esos intercambios publicó el primero de una serie de trabajos sobre la Teoría de la Relatividad, que aparecieron en algunas de las principales revistas científicas de Alemania. Loedel Palumbo se integró plenamente a la vida intelectual de su país adoptivo, llegando a ser considerado como uno de los físicos y filósofos de la ciencia más destacados de la Argentina de su tiempo. Es un ejemplo interesante de la profunda interpenetración de la vida cultural de Uruguay y Argentina en la primera mitad del siglo XX. Compartió también con sus colegas argentinos las consecuencias de vivir en un complejo período histórico.*

PALABRAS CLAVE: *Física en Latinoamérica; relaciones científicas entre Argentina y Uruguay; educación; investigación científica; Einstein y Latinoamérica; teoría de la relatividad; Colegio Libre de Estudios Superiores.*

## INTRODUCCIÓN

La trayectoria del físico uruguayo Enrique Loedel Palumbo (1901-1962) a lo largo del corredor científico «Montevideo-Buenos Aires-La Plata» ejemplifica el intenso intercambio intelectual que tuvo lugar entre Argentina y Uruguay en el primer tercio del siglo XX. Testigos de esa estrecha colaboración, que sin duda ha beneficiado más directamente a la Argentina, han sido figuras destacadas de las ciencias exactas argentinas; entre otros, los uruguayos Marcial R. Candioti (1865-1928), Teobaldo J. Ricaldoni (1861-1923), Félix Cernuschi (1907-1999), Luis Castagnetto y, desde luego, Loedel Palumbo.

En este trabajo primeramente trataremos de presentar el ambiente culto y liberal del Montevideo de principios del siglo XX en el que Loedel Palumbo inició sus primeros estudios. Como alumno del Curso Preparatorio para el ingreso a la Universidad de la

---

[1] *gangui@iafe.uba.ar;*   ORCID iD: https://orcid.org/0000-0002-4864-5348
[2] *e.ortiz@imperial.ac.uk*, ORCID iD: https://orcid.org/0000-0001-8236-8770

República entró en contacto con el Ing. Octavio C. Hansen, que en esos años estaba encargado del dictado de los cursos de física en esa Universidad.

Hansen alentó a Loedel Palumbo a avanzar en sus estudios de física y a publicar su primer libro, en el que da cuenta de sus investigaciones personales en el campo de esa ciencia. Este último es un documento de considerable interés para el estudio de la personalidad científica del autor, que entonces tenía solo 19 años de edad.

En una segunda etapa Loedel Palumbo se trasladó a la Universidad de La Plata, cuyo Instituto de Física había sido generosamente dotado. En La Plata estudió simultáneamente el Doctorado en Física en la Facultad de Ciencias y el Profesorado en Ciencias en la Facultad de Humanidades.

En el Instituto de Física Loedel Palumbo fue discípulo de su director, el conocido físico alemán Richard Gans (1880-1954), de quién recibió una rigurosa formación como investigador científico. En su tesis de doctorado en física, que redactó en 1925, se ocupó de una posible visualización de una molécula orgánica compleja en términos de sus cargas eléctricas y de su comportamiento óptico y magnético. Los resultados de su tesis fueron publicados en la prestigiosa revista alemana *Annalen der Physik,* en 1926.

En marzo-abril de 1925, cuando Albert Einstein visitó la Argentina, Loedel Palumbo no había finalizado aún sus estudios. Sin embargo, la Academia de Ciencias de Buenos Aires lo invitó a contribuir con preguntas específicas en una sesión especial que esa corporación ofrecería al ilustre visitante. Como consecuencia de sus intercambios con Einstein Loedel Palumbo escribió un segundo trabajo que nuevamente publicó en una revista científica alemana: *Physikalische Zeitschrift*.

En la segunda mitad de la década de 1920 Loedel Palumbo había sido incorporado ya al cuerpo docente de la Universidad y de varios institutos de enseñanza secundaria de La Plata. Su trayectoria docente posterior nos permitirá hacer algunas reflexiones acerca del sistema de acumulación de cargos —el llamado *Cumul d'activités*— que era prevalente en esos años en la Argentina. Para lograr reunir un salario adecuado, este sistema obligaba a los investigadores a acumular posiciones de enseñanza en varios institutos; a veces estos estaban asentados en ciudades diferentes de la de su residencia. Hacemos también referencia a los esfuerzos de algunos científicos, que incluye a varios físicos y matemáticos, que bregaron por la introducción del régimen de dedicación exclusiva para los investigadores en actividad.

Finalmente nos ocuparemos de la actuación de Loedel Palumbo en el Colegio Libre de Estudios Superiores, Buenos Aires, a principios de la década de 1930. En un momento en el que la Universidad favorecía, muy particularmente, la formación de *profesionales* el Colegio Libre llegó a convertirse en una de las principales instituciones de apoyo a la cultura superior; en ocasiones se interesó también en la cultura científica.

Conflictos internos en el Instituto de Física, que había pasado ya a manos de científicos locales, sumados a dificultades políticas a nivel nacional, forzaron a Loedel Palumbo a alejarse del grupo dirigente de ese Instituto; este distanciamiento afectó profundamente la dirección de sus actividades. Su doble entrenamiento como físico y como pedagogo le permitió sobrevivir económicamente durante períodos difíciles y, a la vez, contribuir sustancialmente a la actualización y a la mejora del nivel de los textos de enseñanza de las ciencias exactas dirigidos a la juventud. A través de esos textos contribuyó a la modernización de la enseñanza de su disciplina en la Argentina. Sin embargo, a pesar de esas dificultades, Loedel Palumbo continuó sus actividades como investigador, desarrollando una representación que hoy es conocida en el campo de la teoría de la relatividad como los *diagramas de Loedel Palumbo*.

Finalmente, el golpe de Estado militar de 1930 abrió un nuevo período en la Argentina, que no dejó de afectar a actores principales del movimiento científico; ese período excede los límites del presente trabajo.

## LA JUVENTUD DE LOEDEL PALUMBO EN MONTEVIDEO

Enrique Loedel Palumbo nació en Montevideo el 29 de junio de 1901 dentro de una familia uruguaya culta. Sus padres, Juan Edoardo Loedel (1869-1957) y Emilia Palumbo (1869-1962) habían nacido ambos en Montevideo. Un antepasado de su padre fue uno de los principales propulsores de la adopción del sistema métrico decimal en Uruguay. Este fue un elemento importante para la integración de ese país en el concierto internacional[3].

Su madre, maestra y luego directora de una escuela en los alrededores de Montevideo, tuvo considerable influencia en el desarrollo intelectual de sus dos hijos: Enrique y Emilia Zoraida. Loedel Palumbo ha relatado[4] que descubrió su vocación científica y su interés por la física muy tempranamente, hacia los 14 años. También ha hecho referencia al papel que jugó su madre en el desarrollo de esos intereses intelectuales. A partir de aquella edad comenzó a formar en su casa un pequeño laboratorio en el que realizó experiencias científicas, algunas de ellas interesantes, registrando cuidadosamente sus resultados[5].

Luego de terminar su bachillerato de cuatro años cursó los dos años de estudios «Preparatorios» para los estudios de Ingeniería en la Universidad de la República, en Montevideo. Allí trabó contacto con el Ing. Octavio Hansen[6] (1876-1926), quién era entonces profesor de Física en la Facultad de Matemáticas. Hansen es una figura interesante, de la que volveremos a ocuparnos más adelante.

La hermana de Enrique, Emilia Zoraida Loedel Palumbo (1896-1933) mostró también, desde su infancia, un singular talento para las matemáticas; más tarde, fue la primera mujer uruguaya graduada en Ingeniería. Como profesional Emilia desarrolló una intensa actividad en el Ministerio de Obras Públicas de su país y en mérito a sus contribuciones una calle de Montevideo lleva hoy su nombre[7].

## OCTAVIO HANSEN, MAESTRO DE LOEDEL PALUMBO EN EL MONTEVIDEO DE PRINCIPIOS DEL SIGLO XX

El Ing. Hansen es un personaje representativo del mundo cultural y científico de Montevideo en las primeras décadas del siglo XX, que fueron los años de infancia y juventud de Enrique Loedel Palumbo. Nos detendremos, muy brevemente, a considerar algunos aspectos de la personalidad de Hansen[8], y del medio en el que él y sus colegas desarrollaron sus actividades docentes en ese período.

En el último tercio del siglo XIX, el Ateneo del Uruguay, fundado en 1877, continuaba la labor educativa y formativa que había iniciado en Montevideo el Club

---

[3] Loedel, 1864. El trabajo fue publicado el mismo año que otra obra similar Ricaldoni y de la Vega, 1864. Ver detalles de estas obras en Estrada 1912: 292 y 288, respectivamente. El primer autor de la segunda referencia redactó también un tomo breve sobre *Física Popular* Ricaldoni, 1874, una de las primeras obras didácticas sobre ese tema publicadas en el Río de la Plata, mientras que su hijo, el físico Teobaldo J. Ricaldoni, tuvo una considerable influencia en el desarrollo y en la enseñanza de la física en la Argentina.

[4] Loedel Palumbo, 1940.

[5] Xenus, 1920.

[6] Octavio César Hansen, nació en Montevideo el 29 de abril de 1876; su padre, que era también ingeniero, fue Cónsul de Dinamarca en Montevideo.

[7] Loedel Palumbo, 1927; Social Progress, 1928.

[8] Su amigo, el Dr. Agustín H. Mussio, ha dado algunos detalles de su excepcional calidad humana en Mussio, 1926.

Universitario. Este último había ayudado a crear un colegio, que luego fue sostenido por el Ateneo conjuntamente con un grupo de sociedades culturales de Montevideo. Entre las sociedades que cooperaban con el Ateneo se contaba también una Sociedad Científica, contemporánea de su homónima de Buenos Aires[9], y una Sociedad Amigos de la Educación Popular, fundada algo antes, en 1868.

En esos años, el grupo de instituciones de Montevideo mencionado más arriba apoyó el desarrollo de la educación popular, elemental y superior, y también auspició la introducción de capítulos contemporáneos de la ciencia en la enseñanza.

El Ateneo del Uruguay apoyó las corrientes del liberalismo racionalista, que arribaron al Uruguay hacia el último tercio del siglo XIX, en un momento de intenso debate cultural; su tendencia fue claramente laicista. Como ha señalado Alberto Zum Felde[10], con sus trabajos y esfuerzos, el Ateneo y sus sociedades hermanas, contribuyeron a la modernización del pensamiento intelectual del Uruguay, particularmente a partir de la década de 1880.

El Ateneo de Madrid, fundado en 1835[11], fue, sin duda, un modelo cuya influencia sobre el Ateneo del Uruguay, y sobre otras instituciones del Río de la Plata, es innegable. Aquel se autodefinía como una «Sociedad científica, literaria y artística» que se proponía «difundir las ciencias, las letras y las artes por todos los medios adecuados, y favorecer, dentro de su seno, el desarrollo de Agrupaciones que se propongan realizar la investigación científica y el cultivo de las artes y de las letras». Como ocurriría luego con las diversas sociedades rioplatenses, sus actividades culturales estaban abiertas al público educado local.

En 1869 el joven educador José Pedro Varela (1846-1879) fundó en Montevideo una escuela liberal y laica a la que dio el nombre de otro destacado maestro uruguayo, Elbio Fernández (1842-1869), fallecido en plena juventud. Ambos educadores, como antes Domingo F. Sarmiento (1811-1888) en la Argentina, admiraban los avances que la educación popular había alcanzado en los Estados Unidos, particularmente a través de las ideas y de la obra de Horace Mann (1796-1859), y consideraban a Benjamin Franklin (1706-1790) como uno de los principales inspiradores de esa corriente.

Sin duda alguna esos educadores rioplatenses se inspiraban también en las ideas propagadas desde Madrid por la Institución Libre de Enseñanza[12], fundada en 1876 por Francisco Giner de los Ríos (1839-1915) juntamente con un grupo destacado de intelectuales liberales. Su influyente *Boletín* facilitó una difusión amplia de sus ideas, tanto en España como en la América Hispana, donde fue leído con profundo interés.

Hacia 1880 la escuela «Elbio Fernández»[13] contaba con el auspicio del Ateneo del Uruguay y recibía, también, apoyo material de la Sociedad Amigos de la Educación Popular. En esos años había alcanzado ya considerable renombre por la calidad de la enseñanza que impartía y, consecuentemente, por el éxito de sus ex-alumnos. Algunas de las personalidades jóvenes más destacadas de la cultura uruguaya, por ejemplo, el eminente escritor modernista José Enrique Rodó (1871-1917), habían pasado por las aulas de esa escuela.

Octavio Hansen, que también fue uno de sus alumnos, inició sus estudios en un período de auge intelectual del Ateneo. En esos años la escuela «Elbio Fernández» instituyó la costumbre, tomada de Franklin, de que no solamente los profesores, sino también los estudiantes, debían elegir con su voto a quienes consideraban sus mejores compañeros[14]. En el *Escrutinio* de 1885 Hansen fue votado, tanto por su clase como por sus profesores, como el alumno que más se había distinguido por su aplicación al estudio y por su conducta moral[15]. El resto de su vida no haría sino confirmar ese juicio.

---

[9] La Sociedad Científica Argentina, fue fundada en 1872 en Buenos Aires.

[10] Zum Felde, 1987, I: 170-181.

[11] Villacorta Baños, 1978.

[12] Sobre la ILE existe una amplia bibliografía, ver, por ejemplo, el clásico estudio: Jiménez Landi, 1973.

[13] Hoy, Escuela y Liceo Elbio Fernández.

[14] Sociedad de Amigos de la Educación Popular, 1886: 168-169.

Una vez finalizados sus estudios secundarios, Hansen ingresó en la muy recientemente creada Facultad de Matemáticas y Ramas Anexas, dependiente de la Universidad local, donde estudió la nueva carrera de Ingeniería. Una vez graduado adquirió una vasta experiencia profesional: fue el primer director de Obras Municipales y a él se debe el diseño de la Rambla de Montevideo, una de las principales atracciones arquitectónicas de esa ciudad. Además, formó parte del grupo que promovió la creación de la Asociación de Ingenieros del Uruguay, de la que fue elegido su tercer presidente en 1909-1910. Sucedió en ese cargo a un destacado ingeniero, matemático e historiador de la ciencia: el Ing. Eduardo García de Zúñiga (1867-1951).

Hansen pertenece a la primera generación de profesionales universitarios uruguayos que, en números ponderables, comenzaron a hacer entrar a su país en la escena de la técnica contemporánea. Esos jóvenes graduados se esforzaron por actualizar las estructuras tecnológicas del Uruguay intentando, de este modo, reducir la dependencia de su país de los servicios de especialistas extranjeros. Algunos de ellos se asomaron también al mundo de la ciencia pura.

En 1905, sin dejar su carrera profesional en el campo de la ingeniería urbana, Hansen fue designado Profesor de Física y, entre 1909 y 1913, actuó como miembro electo del Consejo Directivo de la Universidad de la República. Justamente en esos años se discutía la reorganización de la Universidad y Hansen jugó, también allí, un papel destacado en los debates promoviendo la modernización de los estudios científicos en la Facultad de Ingeniería. Hasta ese momento la física no había logrado aún ganar un lugar propio dentro de los estudios de la carrera de ingeniería; solo ocupaba una parte muy reducida en el plan de estudios básicos: el llamado *Curso Preparatorio*.

Profundamente interesado en los problemas de la educación, Hansen participó también en la discusión de los nuevos métodos de enseñanza introducidos entre 1904 y 1907, durante el Rectorado de Eduardo Acevedo (1857-1948). Pensaba que el nuevo sistema de promoción sin examen, introducido por Acevedo en la enseñanza secundaria, restaba tiempo a la educación efectiva de los estudiantes. Esto se debía a la necesidad que tenía el profesor de someter a sus alumnos a numerosas pruebas para poder evaluar su nivel de aprendizaje. A la vez, Hansen sostenía que la preparación para la prueba final contribuía a dar al estudiante la posibilidad de «relacionar y comprender cuestiones colocadas á distintas alturas del programa»[16].

Más tarde, también en debates sobre la educación a nivel universitario, Hansen sostuvo que la enseñanza de la física debía «ser ampliada, especialmente en el calor, la electricidad y la meteorología» aconsejando agregar «la acústica para los estudiantes de Arquitectura y la óptica para los de Agrimensura»[17].

Hansen debatió también con Carlos Vaz Ferreira (1872-1958), el humanista más destacado del Uruguay de principios del siglo XX. Uno de sus puntos de divergencia fue, nuevamente, la promoción sin examen, que Vaz Ferreira deseaba extender a los estudios universitarios. Hansen y Vaz Ferreira difirieron también en su apreciación del carácter de los estudios universitarios: mientras que el segundo deseaba dar preferencia a los estudios clásicos dentro de la Universidad, Hansen favorecía prestar una mayor atención a las ciencias. Más tarde, cuando Einstein visitó Montevideo, en 1925[18], la Universidad encargó a Vaz Ferreira, uno de sus más destacados intelectuales, actuar como enlace oficial con Einstein durante esa vista; este último lo recordó con aprecio en el «Diario íntimo»[19] que llevó durante ese viaje.

---

[15] Sociedad de Amigos de la Educación Popular, 1886: 171-172.
[16] Acevedo, 1907: 28-29.
[17] Hansen, 1912: 41.

Este fue el ambiente intelectual que el joven estudiante Loedel Palumbo encontró en Montevideo al ingresar a su Universidad. Una vez en el Preparatorio estableció un contacto estrecho con Hansen, que lo alentó a continuar con sus estudios y, también, a hacer públicos los resultados de sus indagaciones personales. Este fue el origen del primer libro de Loedel Palumbo[20], en el que describió los resultados de sus experimentos.

En esa obra el joven autor dio forma a sus ideas sobre diferentes puntos de la física, ocupándose de las características geométricas de la vena líquida y del estudió de la teoría de los espejos esféricos; para este último tema utilizó ideas básicas de la geometría proyectiva que había aprendido recientemente en el curso Preparatorio. Propuso también una técnica para la determinación de la velocidad de la luz, describió un fotómetro y un telégrafo de su invención y, finalmente, discutió la imposibilidad del movimiento continuo.

Con el objetivo de elevar y modernizar el nivel de la enseñanza Hansen propuso que se ofreciera a los alumnos del Preparatorio la posibilidad de asistir a cursos alternativos al oficial. Sugirió que Loedel Palumbo, a pesar de su extremada juventud y de ser aún un estudiante, fuera encargado de dictar un Curso Libre de Física en el Preparatorio, paralelo al curso oficial, pero más moderno y avanzado.

## LOEDEL PALUMBO Y SUS ESTUDIOS EN LA PLATA: LA INFLUENCIA DE RICHARD GANS

Al finalizar el Curso Preparatorio[21], posiblemente por consejo de Hansen, Loedel Palumbo dejó Montevideo y se inscribió como alumno de la Licenciatura en Ciencias Físico-Matemáticas en la Facultad de Ciencias Físico-Matemáticas de la Universidad Nacional de La Plata (UNLP)[22].

En ese momento esa universidad tenía el laboratorio de física más modernamente dotado de la América Latina. Unos años antes, a poco de que se fundara una Universidad Provincial en la nueva ciudad de La Plata, el Ing. Teobaldo J. Ricaldoni fue designado su primer profesor de física. También se le encargó equipar su laboratorio, para lo que adquirió una colección amplia de instrumentos de física, principalmente de demostración, en una conocida fábrica alemana de instrumentos científicos.

En 1909, en una nueva etapa marcada por la transformación de esa institución Provincial en la nueva *Universidad Nacional de La Plata* (UNLP), se creó en ella un Instituto de Física orientado, muy específicamente, hacia temas de la física moderna. Más tarde, el laboratorio de demostraciones científicas reunido por Ricaldoni fue puesto bajo la dirección del físico alemán Emil Bose (1874-1911), especialmente contratado en Alemania. Bose lo instaló en un edificio nuevo y lo reorientó hacia la enseñanza avanzada y la investigación original en física.

Para alcanzar esos objetivos Bose creó una Licenciatura en Física e inició el dictado de cursos regulares de física a un nivel moderno y avanzado; esos cursos produjeron los primeros graduados en física de la Argentina. En una gran medida, estos últimos dominaron el panorama de esa ciencia en la Argentina por un período prolongado, que se extiende por lo menos hasta mediados del siglo XX.

---

[18] Ortiz y Otero, 2001.
[19] Ortiz, 1995.
[20] Loedel Palumbo, 1920.
[21] Loedel Palumbo pudo haber comenzado a cursar brevemente el primer año de Ingeniería en Montevideo antes de su partida. Xenus, 1920.
[22] Loedel Palumbo, 1940: 2.

Inicialmente Ricaldoni continuó a cargo de los cursos básicos de física. Sin embargo, Bose, que deseaba elevar radicalmente el nivel de los estudios, hizo intentos para modificar esa situación. Lamentablemente, Bose falleció en La Plata en 1911, muy poco después de completar la instalación del nuevo laboratorio de física.

Su sucesor fue otro físico alemán destacado, Richard Gans (1880-1954), que dirigió el Instituto de Física hasta 1925. Gans era un experto de renombre que había contribuido al desarrollo de las teorías modernas del magnetismo; en La Plata contribuyó a formar un grupo amplio de físicos argentinos jóvenes, cuidando de retener a aquellos cuyo entrenamiento había iniciado Bose.

Durante la dirección de Gans algunos de sus alumnos comenzaron a publicar los resultados de sus investigaciones en una revista oficial de la Universidad, fundada por Bose con otro nombre y refundada por Gans en 1913 con el nombre de *Contribución al Estudio de las Ciencias Físicas y Matemáticas*[23]. Algunos de esos estudiantes comenzaron a colaborar también en revistas alemanas de prestigio. Ese fue el singular ambiente de renovación cultural que Loedel Palumbo encontró al iniciar sus estudios en La Plata.

Como hemos indicado ya, simultáneamente con su inscripción en la Licenciatura en Física, en la Facultad de Ciencias Físico-Matemáticas, Loedel Palumbo se registró como estudiante del Profesorado en Matemáticas y Física. Este se cursaba en un ámbito diferente: la Facultad de Humanidades de la misma Universidad. Es posible que este interés por la educación, que como veremos Loedel Palumbo conservó toda su vida, haya sido alentado primeramente por su madre y luego por su maestro Hansen.

En los años en que Loedel Palumbo inició sus estudios en la Facultad de Ciencias los cursos fundamentales de física y matemática estaban a cargo de dos profesores contratados: el antes nombrado Gans y el matemático italiano Hugo Broggi (1880-1965). Estos maestros tenían como asistentes a dos antiguos estudiantes de La Plata: Teófilo Isnardi (1890-1966) y Ramón G. Loyarte (1888-1944). Ambos habían sido iniciados en la investigación por Bose y luego continuaron trabajando bajo la dirección científica de Gans. Tanto Isnardi como Loyarte, lo mismo que un tercer estudiante del grupo inicial, José B. Collo (1897-1968), fueron luego enviados a Alemania para que perfeccionaran su formación científica; Loyarte parece haber sido el único de ellos que obtuvo su doctorado en Alemania[24].

El segundo de los profesores contratados, Broggi, se había doctorado en Göttingen en 1907[25]; su tesis versó sobre la teoría de la probabilidad y su director fue David Hilbert (1862-1943), uno de los matemáticos más originales de su época. Durante su estadía en Argentina Broggi contribuyó, también, a la difusión de resultados recientes de la nueva física cuántica[26]. Hacia fines de la década de 1920 regresó a Italia, donde ocupó una cátedra universitaria; también Gans regresó a su país, en 1925.

Los dos cursos básicos de Física General, que servían para revisar, reforzar y ampliar lo aprendido en la enseñanza secundaria, pasaron de Ricaldoni a Gans, que contaba con la ayuda de Loyarte como asistente. Este último dictó también, en 1923 y 1924, un curso de

---

[23] Llamada también *Contribuciones al Estudio de las Ciencias Físicas y Matemáticas* y, a veces, *Contribución al estudio de las ciencias físicomatemáticas.* La anotaremos como *Contribución* en lo que sigue.

[24] Pyenson y Singh, 1984.

[25] Broggi, 1907.

[26] Gangui y Ortiz, 2009.

Física Especial[27] en el que consideró la mecánica, la conducción del calor y la termodinámica; concluyendo con un panorama de la mecánica estadística y la teoría de los quanta. Sin duda Loyarte tenía un interés serio por la nueva mecánica cuántica y, aparentemente, un buen dominio de su desarrollo contemporáneo; lo mismo puede decirse de la teoría de la relatividad.

Además de los cursos básicos, Gans tenía también a su cargo el curso de Trabajos de Investigación en Física[28], un curso central dentro de esa carrera, que abría horizontes hacia la investigación original en física.

En esos años colaboraba también con el Instituto de La Plata el matemático alemán Paul Frank (1874-1938), que dictaba un curso de Física Matemática. En sus cursos de 1923-1924 (que fueron los que tomó Loedel Palumbo) Frank desarrolló temas de electrodinámica clásica[29].

Frank tenía una buena formación científica, había estudiado en Heidelberg y, en 1899 se doctoró en Leipzig. Según sus biógrafos fue supervisado por el eminente matemático Sophus Lie (1842-1899); posiblemente Lie sugirió el tema y supervisó el comienzo de su tesis, ya que en esos años estaba ya muy enfermo.

Frank se trasladó a la Argentina con un contrato del Instituto Nacional del Profesorado Secundario (INPS), Buenos Aires, junto con otros profesores alemanes de diferentes especialidades. En el INPS, cuya función era formar docentes para el ciclo secundario, Frank dictó cursos de matemáticas superiores. Su colaboración con el Instituto de Física de La Plata se inició en el periodo de Bose, en 1909, y muestra las dificultades que Bose, y luego Gans, experimentaron para atraer, directamente desde Alemania, personal científico calificado.

En 1925 Loyarte se hizo cargo de ese curso en La Plata[30]. En ese mismo período el curso de Físico-química también estaba a cargo de otro ex-alumno del Instituto, Teófilo Isnardi.

Los cursos básicos de análisis matemático estaban a cargo de Broggi y de Emilio Rebuelto; este último era un ingeniero de origen español que tenía una buena formación en la matemática clásica; por muchos años enseñó en Buenos Aires, colaborando a menudo con Julio Rey Pastor (1888-1962) en el dictado de cursos básicos para los estudiantes de ingeniería.

Broggi y Rebuelto dictaban, alternativamente, los dos cursos de análisis matemático siguiendo un texto especialmente redactado para la UNLP por Broggi; esa obra fue publicada entre 1919 y 1927[31]. Finalmente, en 1923 y 1924, Broggi fue encargado también de un curso de Matemáticas Superiores[32] en el que se ocupó, principalmente, de temas de análisis real.

Pareciera que hacia 1925-1926 la Universidad de La Plata intentó ampliar el círculo de intereses dentro de la Facultad de Ciencias, particularmente en dirección al desarrollo de la matemática superior. La incorporación efectiva de Rey Pastor a esa Facultad tuvo lugar precisamente en 1926[33] cuando se le encargó el dictado de un curso avanzado[34]; en esos años Loedel Palumbo había finalizado ya sus estudios.

---

[27] Anuario UNLP, 1923: 44; 1924: 41
[28] Anuario UNLP, 1923: 41; 1924: 41; 1925: 90.
[29] Anuario UNLP, 1923: 36; 1924: 35.
[30] Anuario UNLP, 1925: 84-85. Loyarte cubría temas muy similares a los que antes dictaba en su curso de Física Especial
[31] Broggi, 1919-1927.
[32] Anuario UNLP, 1923: 50; 1924: 46
[33] Ortiz, 2011a.
[34] Anuario UNLP, 1926: 41-42.

## LA EXPLORACIÓN DE LA «FORMA GEOMÉTRICA» DE UNA MOLÉCULA

En diciembre de 1923 Loedel Palumbo se graduó como Profesor de Física en la Facultad de Humanidades, lo que le permitía dictar clases en escuelas de enseñanza media. Como veremos más adelante, esta posibilidad tendría una importancia considerable en su vida profesional, y también personal.

Muy poco más tarde obtuvo su primera posición en la enseñanza: una cátedra de Física en el Colegio Nacional de La Plata, donde inició sus cursos en abril 1924. En el mes de enero de ese año contrajo enlace con María Angélica Gorlero, de 19 años, a quién describiría como su novia desde los 14 años[35].

También en 1924, un año antes de la visita de Einstein a la Argentina, fue designado Conservador en el Gabinete de Física de la Facultad de Ciencias de la Universidad de Buenos Aires[36]. Quizás esta no fuera la posición ideal para un físico teórico, pero es muy posible que el nombre de su designación fuera, simplemente, un requerimiento presupuestario de esa institución. Loedel Palumbo permaneció en ese cargo hasta 1928[37].

A fines de 1925 Loedel Palumbo completó la Licenciatura en Física y, paralelamente con sus estudios del último año de esa carrera, preparó su tesis de doctorado en Física bajo la dirección de Gans, aún director del Instituto de Física y, sin duda, el físico más destacado de la Argentina de esos años.

El 14 de diciembre de 1925 Loedel Palumbo defendió, con éxito, su tesis titulada: «Determinación de las constantes ópticas de la sacarosa por la investigación de la refracción, de la despolarización de la luz de Tyndall y de la polarización circular en soluciones acuosas de tal sustancia»[38]. Poco más tarde la UNLP le otorgó el diploma de Doctor en Física.

En su tesis de doctorado Loedel Palumbo utilizó diversas técnicas ópticas, magnéticas y eléctricas para medir parámetros físicos; entre otros, índices de refracción, de polarización circular y de despolarización de la luz de Tyndall. El tema de sus estudios fue la molécula de sacarosa, y su objetivo intentar atribuir un significado geométrico a las constantes que había medido: deseaba conjeturar acerca de «la forma geométrica» de la molécula de sacarosa. Este era un tema en el que Gans estaba seriamente interesado y sobre el que publicó diversos trabajos en esos años, principalmente, en *Annalen der Physik*[39], pero también en otras revistas científicas alemanas y europeas.

A principios de 1926 Loedel Palumbo presentó un resumen de su tesis para su publicación en *Contribución*, que apareció en ese mismo año[40]. El título de la versión publicada en esa revista es más específico que el de su tesis: "Las constantes ópticas de la molécula de sacarosa. Su *forma geométrica*". En ese trabajo indicó que, a través del estudio de una molécula particular, y utilizando herramientas ópticas, «siguiendo a Gans» pretendía inferir la «forma geométrica» de esa molécula. Concluyó su trabajo expresando que

---

[35] Loedel Palumbo, 1940: 1.
[36] Aunque su nombre oficial era Facultad de Ciencias Exactas, Físicas y Naturales, en esos años era conocida como Facultad de Ingeniería, lo que reflejaba la centralidad de los estudios de ingeniería en esa institución.
[37] Loedel Palumbo, 1940: 2.
[38] Este es el título que indica Westerkamp, 1975: 97, en su detallada lista de tesis de doctorados en Física en las diversas universidades argentinas.
[39] Gans, 1921.
[40] Loedel Palumbo, 1926a.

«debemos imaginar la molécula de sacarosa, en su comportamiento óptico, como algo así como un elipsoide torcido helicoidalmente, para explicar con ello la polarización circular»[41].

En su publicación, Loedel Palumbo agradeció a Gans, no solamente por la dirección de su trabajo, sino también por haber «recibido siempre [de él] inapreciables consejos y múltiples atenciones». En esos años Gans estaba haciendo un esfuerzo serio por educar a sus alumnos, no solamente a hacer investigaciones originales, sino también a concretar sus resultados, redactarlos y presentarlos en una forma publicable, tanto en el país como en la prensa científica internacional. Ese fue el caso de la tesis de doctorado de Loedel Palumbo: una versión más reducida, con un apéndice teórico de Gans, fue publicada en *Annalen der Physik*[42].

En el mismo año, 1926, J. Ewles, del Laboratorio de Física de la Universidad de Leeds, comentó los resultados del trabajo de Loedel Palumbo en *Physics Abstracts*. Decía Ewles que los datos presentados por Loedel Palumbo «contribuirían a aclarar la forma en la que la molécula de sacarosa está construida, con cargas positivas y negativas»[43].

Pero un año antes, en 1925, Gans dejó la Universidad de La Plata para hacerse cargo de una cátedra en Königsberg, de modo que Loedel Palumbo fue su último discípulo en La Plata. Como regresaba a Alemania antes de que su alumno pudiera presentar su tesis, ya que aún estaba cursando el último año de su carrera y, primeramente, debía rendir sus exámenes finales, Gans dejó documentado que esa tesis contaba con su total aprobación. Sin embargo, su propuesta encontró ciertos reparos formales dentro de la Facultad, que el Consejo Universitario prontamente removió[44].

Este y otros documentos del Archivo de esa Universidad[45], lo mismo que la correspondencia de Emil Bose y Margrete Heiberg de Bose (1865-1952)[46], indican que luego de la partida de Gans surgieron diferencias serias entre algunos miembros del Instituto de Física, que antes habían sido sus alumnos. En aquella correspondencia es posible detectar esas diferencias aun inmediatamente después del fallecimiento de Bose.

Sin duda Gans debe haber tenido motivos serios para dejar La Plata, ya que en ese período las condiciones de vida en Alemania eran, todavía, sumamente duras a causa de las secuelas de la Primera Guerra Mundial, incluso para un profesor universitario. Precisamente alrededor de esos años, en 1923-1924, se registró el pico histórico máximo de emigración en el sentido contrario: de Alemania hacia la Argentina[47].

Años más tarde, una vez finalizada la Segunda Guerra Mundial, en un homenaje tributado a Gans en celebración de sus 70 años y de su regreso a la Argentina, su ex-alumno Enrique Gaviola (1900-1989) declaró: «Circunstancias lamentables para la física argentina alejaron a Gans de La Plata entre 1925 y 1947. Sus viejos alumnos nos sentimos regocijados por su vuelta»[48].

**LOEDEL PALUMBO Y LA VISITA DE EINSTEIN A LA ARGENTINA, EN 1925**

Precisamente en 1925, Albert Einstein (1879-1955) visitó la Argentina invitado a dictar una serie de conferencias sobre su teoría de la relatividad en la Universidad de Buenos Aires.

---

[41] Loedel Palumbo, 1926a: 78.
[42] Loedel Palumbo, 1926b.
[43] Ewles, 1926.
[44] Archivo UNLP, 1926a
[45] Por ejemplo, los documentos Archivo UNLP, 1926b y 1926c.
[46] Ortiz, 2021.
[47] Rinke, 2005: 30.
[48] Gaviola, 1950

Además, dictó conferencias especiales en las universidades de La Plata y Córdoba y en instituciones de la comunidad judía de la Argentina[49].

Con motivo de la visita de Einstein, la recientemente establecida Academia Nacional de Ciencias Exactas, Físicas y Naturales de Buenos Aires decidió organizar una reunión especial sobre la teoría de la relatividad. Esa tertulia tenía un doble carácter: primeramente, ofrecer un homenaje especial al visitante, al que se le entregaría el diploma de Académico Honorario y, también, brindar a los expertos argentinos una oportunidad de interactuar con el sabio alemán.

Esa sesión académica especial tuvo lugar el 16 de abril y fue coordinada por su entonces presidente, el naturalista Dr. Eduardo L. Holmberg (1852-1937), uno de los científicos más sobresalientes de la Argentina de esa época. Para la realización de esa reunión Holmberg contó con la colaboración del Ing. Nicolás Besio Moreno (1879-1962)[50]. Besio Moreno, que había sido director del Observatorio y decano de la Facultad de Ciencias Físico-Matemáticas de la Universidad de La Plata fue, más tarde, autor de contribuciones de considerable interés para la historia de la matemática en la Argentina.

La Academia invitó a aquella reunión a varios Académicos, y también a especialistas en temas avanzados de física, ofreciéndoles la oportunidad de proponer al homenajeado preguntas sobre su teoría. Entre los científicos invitados figuraban los Académicos Dr. Ramón G. Loyarte y Dr. Horacio Damianovich (1883-1959) y tres jóvenes investigadores: los físicos Dr. Teófilo Isnardi y Dr. José B. Collo y el astrónomo Ing. Félix Aguilar (1884-1943). Los tres últimos, que más tarde serían también miembros de esa Academia, acababan de publicar una exposición breve sobre la teoría de la relatividad y sus implicaciones astronómicas[51]. Ese folleto fue el resultado de un ciclo de introducción a la teoría de la relatividad organizado por la Sociedad Científica Argentina en 1922-1923, en preparación a la visita de Einstein[52].

También fue invitado a esa sesión, para contribuir con sus preguntas, el joven estudiante Enrique Loedel Palumbo, que no solo no había concluido aún su doctorado, sino que ni siquiera había finalizado aún sus estudios de la Licenciatura en Física en la Universidad de La Plata.

Tanto las preguntas formuladas a Einstein, como sus respuestas, fueron publicadas en los *Anales* de esa Academia[53], que entonces estaban incorporados a los *Anales de la Sociedad Científica Argentina*. Aunque las preguntas de Aguilar, Collo, Damianovich, Isnardi y Loyarte revelan conocimiento de esa teoría, las de Loedel Palumbo, muy específicas, sugieren un interlocutor que en ese momento se encontraba tratando de investigar en el campo de la teoría de la relatividad.

En otra parte, hemos indicado que durante algunos de sus viajes Einstein llevó un «Diario íntimo»[54], en el que volcó sus impresiones. Si bien ese Diario no puede tomarse como un reflejo totalmente fiel de sus actividades, ya que en él hay omisiones, agregados y nombres supuestos, sus impresiones sobre esa reunión no fueron halagadoras. El mismo 16

---

[49] Ortiz, 1995. Gangui y Ortiz, 2005, 2008, 2009 y 2014.

[50] Academia, 1929: 337-347. Ver también Deulofeu, Galloni y Santaló, 1975: 187-188.

[51] Collo, Isnardi, y Aguilar, 1923-1924.

[52] El contenido técnico de ese trabajo ha sido analizado en Gangui y Ortiz, 2011.

[53] Academia, 1929.

[54] Detalles del contenido de ese *Diario* fueron publicados en Ortiz, 1995. Su autor agradece al Prof. Gerald Holton (Harvard University) haberle facilitado acceso a ese material que, hasta esa fecha, había permanecido como un documento reservado

de abril de 1925 Einstein escribía en la página 23 de su Diario: «A uno le preguntan cuestiones [insertado: científicas] muy tontas, de modo que fue difícil permanecer serio»[55].

Si las minutas oficiales de esa reunión, incompletas, pudieran tomarse como una referencia confiable, Einstein habría encontrado algo más interesante la pregunta que le formuló el joven Loedel Palumbo, que hizo que el visitante se detuviera a responderla, revelando cierto interés.

La pregunta de Loedel Palumbo fue la siguiente: «¿es posible hallar una representación de la superficie espacio-tiempo de dos dimensiones en un espacio euclídeo de tres?»[56]. El problema, supuestamente, había conducido a ese joven físico a formular un sistema de ecuaciones diferenciales parciales no-lineales que aún no había logrado resolver. Según esa misma publicación, «el profesor Einstein contesta que [esas ecuaciones] no han sido resueltas, y que el problema de investigar la forma de la superficie espacio-tiempo sería muy interesante»[57].

Podría decirse que Loedel Palumbo había pasado de tratar de inferir la forma geométrica de la molécula de sacarosa en tres dimensiones a preguntarse acerca de la forma geométrica de la superficie espacio-tiempo de dos dimensiones en un espacio euclídeo de tres.

Un año más tarde, a principios de 1926, Loedel Palumbo consiguió resolver el problema que había planteado a Einstein y publicó su solución, primero en *Physikalische Zeitschrift*[58], donde lo presentó el 17 de abril, y luego en idioma español en *Contribución*[59], donde lo entregó para su publicación un mes más tarde, el 18 de mayo. Los dos textos son aproximadamente los mismos, pero hay ligeras diferencias.

El trabajo de Loedel Palumbo es una nota breve, no ya un trabajo fundamental[60], pero el problema que resuelve no era en ese momento necesariamente trivial. En su referencia sobre ese trabajo, publicada en el *Jahrbuch der Mathematik*, el Prof. Friedrich Möglich (1902-1957), de Berlín, consigna, en una línea, que el problema ha sido abordado y resuelto por el autor[61].

Por su parte, el físico británico J. S. G. Thomas, de Londres, en una referencia ligeramente más sustancial publicada en *Physics Abstracts* señaló que la nota de Loedel Palumbo «ilustra [sobre] la significación de la teoría general de la relatividad»[62], y que lo hace reduciendo el elemento de línea de una forma cuatridimensional al de una forma espacial euclídea en tres dimensiones, en el caso especial en el que $\vartheta$ = constante y $\phi$ = constante[63].

Si hemos de tomar como fidedigna la versión de los *Anales de la Academia*, dudosa en más de un aspecto, Einstein, Loedel Palumbo, los editores de *Contribución* (que publicaron su nota en Argentina) y los editores de *Physikalische Zeitschrift* (que publicaron su nota en Alemania), ninguno de ellos parece haber estado al tanto de que ese problema había sido ya considerado, cuatro años antes de aquella reunión, y también resuelto con una mayor generalidad. Así lo ha mostrado el Prof. Jean Eisenstaedt en su interesante trabajo[64], donde destaca que el autor de esa solución fue el geómetra Edward Kasner (1878-1955), de la Universidad de Columbia[65].

---

[55] Ortiz, 1995: 114.

[56] Academia, 1929: 346-347.

[57] Academia, 1929: 347.

[58] Loedel Palumbo, 1926d.

[59] Loedel Palumbo, 1926c.

[60] El Dr. Einsenstaedt, Laboratoire de Gravitation et Cosmologie, Université Pierre et Marie Curie, Paris, nos ha escrito (08.02.1999) que considera que «Il s'agit en somme de un travail tout à fait correct pour l'époque mais sans grande originalité».

[61] El Dr. Möglich dice, simplemente, de este trabajo: «Se calcula el campo de una masa puntiforme y se ilustra con figuras». Möglich, 1926.

[62] «The significance of gravitation in Einstein's general theory of relativity is illustrated by reduction of the four-dimensional form of the linear element to a three dimensional Euclidean space form in the special case in which $\vartheta$ = const and $\phi$ = const».

[63] Thomas, 1926.

En una nota breve, publicada en *Tohoku Mathematical Journal*, el matemático japonés Uichirô Azuma, de Sendai, hizo una reseña breve del trabajo de Loedel Palumbo[66], indicando que en la primera parte el autor había elegido la coordenada de tiempo como un número imaginario, con lo que también lo era la superficie. En la segunda parte Loedel Palumbo eligió la coordenada z como un número imaginario, con lo que el tiempo resultaba real. Esencialmente, este autor señalaba la habilidad de Loedel Palumbo para razonar geométricamente que, más tarde, lo llevaría a una nueva e interesante contribución a la teoría de la relatividad: los llamados *Diagramas de Loedel Palumbo*, de los que nos ocupamos separadamente.

Sin exagerar su importancia, podría decirse que aquella nota breve de Loedel Palumbo, de 1926, abrió dos capítulos nuevos para la física en la Argentina: el del intento de hacer investigaciones originales en física teórica y el de la teoría de la relatividad como tema de investigación. Es decir, no ya de divulgación popular o avanzada, como había sido el caso hasta entonces.

En junio del mismo año Loedel Palumbo volvió a ocuparse de temas de relatividad en un trabajo sobre la velocidad de la luz en un campo gravitacional[67]. Sin embargo, después de la partida de Einstein de la Argentina el interés por la teoría de la relatividad se fue extinguiendo, gradualmente, quedando reducido a las tareas de muy pocos cultores de la física y, casi exclusivamente, con propósitos didácticos. Como hemos señalado en otra parte[68], en esos años, tanto en la Argentina como fuera de ella los temas de la nueva mecánica cuántica fueron absorbiendo el interés de los investigadores en física teórica, desacelerando el antiguo interés por la relatividad.

Efectivamente, en 1927, solo dos años después de la visita de Einstein, Loyarte publicó en *Contribución* un trabajo titulado "La nueva mecánica atómica"[69]. En esa nota se ocupó de los avances que se habían realizado recientemente en el campo de la mecánica cuántica, describiendo las sucesivas etapas que habían conducido al desarrollo de esa *nueva mecánica*, como señala A. Daniell en la reseña de ese trabajo que escribió para *Physics Abstracts*[70].

También en ese mismo año Loyarte dictó un curso avanzado de mecánica cuántica en su cátedra de Física Teórica en la Universidad de La Plata. Dos de sus ex-alumnos, profesores ya en esa Universidad: el matemático Alberto E. Sagastume Berra (1905-1960) y el físico Rafael Grinfeld (1902-1969), tomaron notas de las clases de Loyarte y con base en ellas comenzaron a serializar una monografía extensa sobre la nueva "mecánica atómica" en los *Anales de la Sociedad Científica Argentina*[71]. Al comienzo de su monografía esos autores indicaron claramente que su trabajo era, en verdad, un apunte «re-trabajado» de las notas de clase de Loyarte. Más tarde, la Universidad de la Plata publicó esa monografía en forma de libro[72]. Es posible que esas notas hayan sido la primera exposición sistemática avanzada de la mecánica cuántica realizada por autores locales en nuestra lengua.

---

[64] Eisenstaedt, 1989: 218 y 233.

[65] Kasner, 1921a. Ver también Kasner, 1921b y 1921c. Más tarde, Kasner contribuyó repetidamente con sus notas a la *Revista de la Unión Matemática Argentina*. Su famoso libro Kasner y Newman, 1940, fue reseñado por el destacado escritor argentino Jorge Luis Borges en la *Revista Sur*, Buenos Aires. Sin duda, tuvo un impacto considerable sobre sus concepciones del uso de ideas de la matemática en la literatura, Ortiz, 1998.

[66] Azuma, 1928.

[67] Loedel Palumbo, 1926e.

[68] Ortiz, 2011b: 25-32.

[69] Loyarte, 1927.

[70] Daniell, 1928.

## LOEDEL PALUMBO Y LA POLÍTICA CIENTÍFICA DEL *CUMUL D'ACTIVITÉS* FRANCÉS EN LA ARGENTINA

Luego de su primera designación en La Plata, como profesor de física en el Colegio Nacional de La Plata en abril 1924, y en Buenos Aires, como Conservador en el Gabinete de Física de la Facultad de Ciencias, a las que hemos hecho referencia más atrás, Loedel Palumbo fue designado Jefe de Trabajos Prácticos de Física en la Facultad de Ciencias Físico-Matemáticas en 1926. El año siguiente adquirió una segunda cátedra en el Colegio Nacional de La Plata, y fue designado profesor suplente de Física General (cursos A o B) en la Facultad de Ciencias Físico-Matemáticas. A la vez, fue designado profesor suplente de Geografía Matemática en la Facultad de Humanidades. En la década siguiente, en 1936, fue designado profesor de física en el Liceo de Señoritas[73], también en La Plata.

Sorprende quizás hoy, aunque no en su época, la diversidad de cargos que ocupaba este destacado científico para componer un salario que asegurara su subsistencia y la de su familia inmediata.

A pesar de la prédica de algunos investigadores destacados, como el fisiólogo Bernardo A. Houssay (1887-1971), el matemático Rey Pastor y, más tarde, también el físico Gaviola junto con otros científicos, el *Cumul d'activités* francés —el empleo múltiple— continuó siendo una necesidad absoluta para los investigadores científicos argentinos. También lo fue para los investigadores que trabajaban en otras áreas de la cultura.

La toma de conciencia de que la investigación científica debía ser considerada como una actividad profesional exclusiva fue un proceso que tuvo una evolución lenta y compleja en la Argentina. Las ideas presentadas por Santiago Ramón y Cajal (1852-1934) en su importante libro acerca de la investigación científica[74] fueron un ingrediente significativo y temprano en ese proceso; su impacto en Argentina no puede desestimarse.

Por otra parte, las visitas extendidas de científicos e intelectuales españoles a la Universidad de Buenos Aires, uno de los cuales fue Rey Pastor, formados con ayuda de la Junta para Ampliación de Estudios e Investigaciones Científicas, fueron posibles —y extremadamente exitosos— gracias a la conjugación de dos elementos. Por una parte, la filantropía de la Institución Cultural Española en Buenos Aires, por otra, el apoyo que la Junta, presidida entonces por Ramón y Cajal, prestó desde Madrid en la selección de los profesores visitantes. A través del ejemplo de la Junta para Ampliación de Estudios, esas visitas dieron una evidencia clara de que el apoyo oficial a la investigación científica podía rendir resultados tangibles a corto plazo.

En diversas fuentes argentinas en las que se exploran ideas modernas acerca de la investigación científica, por ejemplo, en el discurso de incorporación a la Academia de Ciencias de Buenos Aires que pronunció Besio Moreno en 1923[75] es posible rescatar reflexiones directamente inspiradas en la obra de Ramón y Cajal que mencionamos más atrás. También mostraron esas experiencias extranjeras que, en la Argentina, era necesario repensar la política científica que estaba implícita en el *empleo múltiple* al estilo francés. Contemporáneamente, Loyarte, en su discurso de enero de 1928, leído al hacerse cargo de la presidencia de la Universidad de La Plata, al que aludimos más abajo, hizo presente la necesidad de un cambio fundamental en la política de apoyo a la investigación científica en la Argentina.

---

[71] Sagastume Berra y Grinfeld, 1928.
[72] Sagastume Berra y Grinfeld, 1930.
[73] Loedel Palumbo, 1940: 2.
[74] Ramón y Cajal, 1899.
[75] Besio Moreno, 1928.

Lo mismo puede decirse de los esfuerzos en favor de la creación de becas internas realizados inicialmente en el ámbito de la Sociedad Científica Argentina y, luego, por iniciativa de Holmberg, desde la remodelada Academia de Ciencias de Buenos Aires.

Más tarde esos esfuerzos fueron retomados por la Asociación Argentina para el Progreso de las Ciencias, que se vio obligada a recurrir a la cooperación financiera de la Fundación Rockefeller de los Estados Unidos[76]. Sin embargo, el camino fue largo: las designaciones a tiempo completo para personal docente en los diferentes niveles académicos solo comenzaron a cambiar sustancialmente en la Universidad de Buenos Aires, y esto recién ocurrió hacia fines de la década de 1950[77].

**LAS VISITAS EUROPEAS DE LOEDEL PALUMBO**

Desde la partida de Gans Loyarte quedó a cargo de la dirección del Instituto de Física; sin embargo, dos años después, en diciembre de 1927, fue elegido presidente de la Universidad de La Plata: la misma institución a la que unas dos décadas más atrás había ingresado como alumno. El período de su designación cubría los años 1928 a 1930.

En su discurso inaugural, como presidente de la Universidad, Loyarte se ocupó de otro aspecto importante para el entrenamiento de científicos jóvenes: las becas externas. Desde su nueva posición promovió el «envío sistemático de egresados distinguidos a los grandes institutos del extranjero»[78], como años atrás había sido su propio caso por una inteligente sugerencia de su maestro Bose.

Durante el ejercicio de la presidencia de la Universidad Loyarte logró instaurar esa práctica y, como consecuencia de su nueva política cultural, en 1928 Loedel Palumbo obtuvo una beca de la UNLP para continuar sus estudios en el extranjero.

Con esa beca se trasladó a Berlín[79], en cuya universidad realizó estudios de física bajo la dirección de dos distinguidos maestros: Max Planck (1858-1947) y Erwin Schrödinger (1887-1961). Asimismo, asistió a los seminarios de Max von Laue (1871-1960) y de Wilhelm Nernst (1864-1941). Loedel Palumbo ha indicado que en ese mismo período estuvo también en contacto con Einstein[80], lo que no es sorprendente teniendo en cuenta sus contactos anteriores, y con Hans Reichenbach (1891-1953) que en esos años trataba de desarrollar una fundamentación más precisa de las ciencias exactas.

A través de Reichenbach, Loedel Palumbo tomó un contacto directo con el movimiento positivista lógico en su núcleo local: el llamado *Círculo de Berlín*[81]. Más tarde señaló su afinidad por ese movimiento y por sus ideas. Decía Loedel Palumbo en 1940 que:

---

[76] Ortiz, 2015.

[77] La Universidad de Buenos Aires solo introdujo la dedicación exclusiva a nivel de Jefes de Trabajos Prácticos en 1958.

[78] *El Argentino*, 1927.

[79] Loedel Palumbo, 1959: 3.

[80] Loedel Palumbo, 1940: 2.

[81] *Die Berliner Gruppe*; existía también un círculo paralelo en Viena (*Der Wiener Kreis*), con algunas diferencias de interpretación.

«En algunas épocas de mi vida he estudiado filosofía con fervor, en mi juventud fui racionalista cartesiano, más tarde escéptico y por el momento el empirismo consecuente del Círculo de Viena[82] cuenta con mis simpatías»[83]. En las décadas de 1930 y 1940 los estudios de Loedel Palumbo reflejan un interés profundo por la discusión de la causalidad y de los fundamentos de las ciencias físicas, temas a los que contribuyó con varios aportes que indicaremos más adelante. Más tarde otro graduado de La Plata, Mario Bunge[84] retomaría el tema de la causalidad en las ciencias físicas[85].

Aunque sus intereses lo identifican como un físico teórico, durante su estadía en Berlín hizo esfuerzos serios por perfeccionar y actualizar su técnica experimental. Para ello asistió a los cursos de física experimental que dictaba Peter Pringsheim (1881-1963), reciente director de la tesis de doctorado de Gaviola, en el Walther Nernst-Institut für Physikalische und Theoretische Chemie de esa misma universidad. En diciembre de 1929, ya de regreso en La Plata, Loedel Palumbo comenzó a utilizar su entrenamiento experimental para elevar el nivel de las prácticas de laboratorio de física en los colegios de La Plata. Su expediente personal en el Liceo de Señoritas hace referencia a sus donaciones de instrumentos de laboratorio y, en una nota oficial, se le agradece también su contribución al desarrollo del laboratorio de ese Liceo.

A principios de la década de 1930 Loedel Palumbo continuó profundizando sus intereses en la filosofía de la ciencia y en 1935 volvió a viajar a Europa, esta vez para participar en la primera reunión del *Congrès International de Philosophie Scientifique*[86]. A esa reunión, que se realizó en la Sorbonne con la participación de diversas personalidades[87], también asistió Reichenbach, que en esa época había dejado ya su país y se había refugiado en Turquía.

## LOEDEL PALUMBO EN EL COLEGIO LIBRE, A PRINCIPIOS DE LA DÉCADA DE 1930

Entre sus dos viajes a Europa, Loedel Palumbo continuó en contacto profesional con las Facultades de Humanidades y de Ciencias de la UNLP pero, según su documentación en el Archivo de la UNLP, al dejar la Argentina interrumpió su colaboración docente con la Universidad de Buenos Aires, la que no parece haberse reanudado a su regreso. En ese momento, esa colaboración hubiera sido oportuna para reforzar el desarrollo de la moderna física teórica en Buenos Aires.

Sin embargo, Loedel Palumbo no cortó sus contactos con Buenos Aires. Por una parte, el Instituto Nacional del Profesorado Secundario lo designó profesor de Mecánica en 1932, y luego de Física Matemática en 1933. Recordemos que una década más atrás, como alumno de la UNLP, Loedel Palumbo había asistido a las clases de Física Matemática que dictaba en La Plata el físico alemán Paul Frank, contratado por el Instituto del Profesorado y luego compartido con la Universidad de La Plata como profesor. La situación se había

---

[82] Sobre el Círculo de Viena ver Schmitz, 2009: 104, en donde hay una breve referencia a la reunión de Paris de 1935.
[83] Loedel Palumbo, 1940: 1.
[84] Bunge, 1959.
[85] Ortiz, 2019.
[86] Premier Congrès, 1936.
[87] Entre ellos, Alfred Ayer (1910-1989), Rudolf Carnap (1891-1970), Federigo Enriques (1871-1946), Philipp Frank (1884-1966), Albert Lautman (1908-1944) y Alfred Tarski (1901-1983).

revertido mostrando, de alguna manera, el éxito del sistema de incorporar profesores extranjeros destacados capaces de formar discípulos locales

También en esa década, un grupo importante de intelectuales comenzó a criticar, más persuasivamente que en el pasado, la excesiva concentración del esfuerzo de las universidades en la formación de profesionales. Esos críticos no dudaban de la importancia de la formación de profesionales eficientes para el desarrollo de la Nación, pero censuraban la casi total ausencia de apoyo oficial a los centros que intentaban promover la investigación científica original. Como hemos indicado más atrás en algunos círculos esa actividad comenzaba a ser reconocida como indispensable para crear una cultura científica sólida en el país. La polémica sobre la necesidad de la investigación científica no se limitó a los centros universitarios o a la Academia de Ciencias de Buenos Aires: alcanzó también a sectores técnico-científicos de las fuerzas armadas[88].

Una de las instituciones independientes, de alto nivel cultural, que trataron de ocupar ese espacio fue el Colegio Libre de Estudios Superiores. Fundado en Buenos Aires en mayo de 1930, pocos meses antes del golpe militar de septiembre de ese año, representaba una alternativa frente a la carencia, o a la fragilidad de los institutos de altos estudios de la Universidad. En un grupo de áreas de la cultura el Colegio Libre se convertiría en una de las tribunas más altas y efectivas para la promoción de estudios avanzados. Su Acta de Fundación establecía que sería «un organismo exento de carácter profesional, destinado a contribuir al desarrollo de los estudios superiores» y, luego, que no intentaba ser «ni Universidad Profesional, ni tribuna de vulgarización»[89].

Aunque no formó parte de su grupo fundador, uno de los inspiradores intelectuales del Colegio Libre fue el fisiólogo alemán Georg Friedrich Nicolai (1874-1964), ex-profesor de Fisiología en la Universidad de Berlín y co-firmante, con Einstein, del documento pacifista de 1914. Durante la Primera Guerra Nicolai escribió un famoso libro, *Die Biologie des Krieges*, en el que analizaba críticamente la guerra como un fenómeno biológico[90]. Durante esa guerra su obra fue prohibida en Alemania pero, impresa en Suiza, fue introducida ilegalmente en su país. Hacia el final de la guerra Nicolai protagonizó una dramática evasión desde Alemania y, una vez concluida esa guerra, regresó a su país y fue restituido como Profesor en Berlín; a partir de entonces su obra de 1917 tuvo una difusión considerable.

Sin embargo, desde principios de la década de 1920 Nicolai experimentó dificultades debido a su antigua posición pacifista: estudiantes de la extrema derecha sistemáticamente interrumpían y creaban disturbios en sus clases. Si bien estos no contaban con el beneplácito oficial de las autoridades universitarias, contaban sí con su indolencia.

Poco después de la Reforma Universitaria de 1918, la Universidad de Córdoba lo invitó a incorporarse a ella como profesor de fisiología, lo que Nicolai aceptó como una medida temporaria que luego se convertiría en definitiva. En Córdoba volvió a encontrarse con Einstein, cuando este visitó la Argentina, en 1925. Más tarde se trasladó a Santiago de Chile, donde falleció en 1964.

Es interesante señalar que la primera obra que publicó el sello editorial del Colegio Libre (Editorial CLES), fue precisamente una traducción al español de la *Biología de la Guerra*, en la que incluyó el prólogo que Romain Rolland (1866-1944) había escrito para esa obra[91].

---

[88] Ortiz, 2015.
[89] Colegio Libre de Estudios Superiores, 1946: 2.
[90] Nicolai, 1917.
[91] Nicolai, 1932.

A principios de la década de 1930, y luego en 1931 y 1932, Nicolai colaboró con el Colegio Libre[92] dictando conferencias sobre la psicología, sobre la influencia del medio geográfico sobre la humanidad, y sobre su visión de los cambios ocurridos en Rusia a partir de la Revolución de 1917.

Loedel Palumbo fue también invitado a colaborar con el Colegio Libre, donde dictó en 1931 un ciclo de nueve conferencias sobre estudios recientes acerca de la posible estructura del átomo. El año siguiente el Colegio Libre lo invitó nuevamente, y en ese segundo ciclo dictó seis conferencias sobre el núcleo atómico. Más tarde, *Cursos y Conferencias*, la revista oficial del Colegio Libre, publicó resúmenes de sus conferencias sobre la estructura del átomo[93] y, tres años después, publicó un estudio sobre el problema de la causalidad en la física[94].

En esos años el tema del segundo grupo de conferencias de Loedel Palumbo: el núcleo atómico, era de gran actualidad. Poco después de las últimas conferencias de Loedel Palumbo, en julio-agosto de 1934, Enrico Fermi[95] visitó las universidades de Buenos Aires y Córdoba con el patrocinio del Instituto Italiano de Cultura (otro émulo local de la Institución Cultural Española de Buenos Aires). En esa ocasión dictó un curso sobre la nueva física en la Facultad de Ciencias. En sus conferencias III, IV y V discutió problemas contemporáneos mostrando que el foco de la investigación en física moderna se había desplazado hacia el estudio de la estructura del *núcleo* atómico, utilizando la desintegración artificial como una herramienta de análisis.

En ese mismo año el Colegio Libre invitó también a Loyarte; este último dictó dos «conversaciones» sobre las relaciones entre la mecánica clásica y la mecánica cuántica[96] que, lamentablemente, no fueron recogidas en una publicación; tampoco existe un resumen de esas conferencias.

**CONSIDERACIONES FINALES**

Loedel Palumbo inició sus estudios en Montevideo, en un ambiente cultural elevado; allí fue alumno de Octavio Hansen, uno de los primeros profesores de física de la Universidad de la República. Hansen había sido formado en la escuela «Elbio Fernández» dentro de las tradiciones liberales y laicas que entonces predominaban en los sectores intelectuales de esa ciudad; luego estudió ingeniería en la Universidad de la República.

A principios del siglo XX el Instituto de Física de Universidad de La Plata se convirtió en un significativo polo de atracción para los jóvenes de ambas orillas del Plata, interesados en las ciencias físicas. Formados por su entonces director, el profesor alemán Richard Gans, algunos de esos jóvenes alcanzaron cierto relieve en el campo de la física.

Entre ellos se destacó Enrique Loedel Palumbo, que inició sus estudios en Uruguay y luego se trasladó a La Plata. En el campo de la investigación llegó a ser uno de los discípulos más brillantes entre los alumnos que Gans formó durante su estadía en la Universidad de esa ciudad. Durante la visita de Einstein a la Argentina Loedel Palumbo fue, sin duda, el interlocutor local más interesante. Loedel Palumbo se integró también a algunos de los principales círculos intelectuales de su época: en Buenos Aires participó en las actividades del Colegio Libre de Estudios Superiores.

---

[92] Colegio Libre de Estudios Superiores, 1946: 56.
[93] Loedel Palumbo, 1931b.
[94] Loedel Palumbo, 1934.
[95] Fermi, 1934.
[96] Colegio Libre de Estudios Superiores, 1946: 54.

Como el resto de sus colegas compartió la investigación científica con diversas tareas de enseñanza. En un segundo período de su vida, que no consideramos en este trabajo y que, en parte, fue cruzado por su separación de la Universidad de La Plata por razones políticas, se destacó por la redacción de obras de texto modernas que dieron un auxilio valioso a la enseñanza de la física en nuestra lengua. Es posible que esos textos, calificados y modernos, producidos por expertos graduados en el Instituto de Física de La Plata constituyan una de las contribuciones sociales más visibles del experimento de crear centros de alta excelencia científica promovido por los fundadores de la Universidad de La Plata.

**AGRADECIMIENTOS**

**FUENTES**

Anuarios de la Universidad Nacional de La Plata,
    *Anuario para el Año 1923*, 13, No. 57.
    *Anuario para el Año 1924*, 14, No. 60.
    *Anuario para el Año 1925*, 15, No. 62.
    *Anuario para el Año 1926*, 16, No. 70.

Archivo de la Universidad Nacional de La Plata,
Expedientes,
    *(1926a) No. 1026, Decano Julio Castiñeiras a Dr. Teófilo Isnardi, 11 enero 1926.*
    *(1926b) No. 2867, Dr. Ramón G. Loyarte a Decano Julio Castiñeiras, 21 mayo 1926.*
    *(1926c) No. 2172, Decano Julio Castiñeiras a Dr. Ramón G. Loyarte, 24 noviembre 1926.*

Ficha Bio-Bibliográfica:
    (1940). Loedel Palumbo, Enrique, junio 25, 1940. La Plata: Colegio Secundario de Señoritas.

Antecedentes, científicos y docentes:
    (1959). Loedel Palumbo, Enrique. La Plata: Universidad Nacional de La Plata.

**BIBLIOGRAFÍA**

Academia, "Academia: Recepciones y distinciones. Recepción del doctor Alberto Einstein en la Sesión Especial de la Academia el día 16 de abril de 1925". *Anales de la Academia Nacional de Ciencias Exactas, Físicas y Naturales de Buenos Aires*, I, (Buenos Aires, 1929):

337-347. [Originalmente publicado en *Anales de la Sociedad Científica Argentina*, 107(1): 337-347].

Acevedo, Eduardo, *La enseñanza universitaria en 1906. Informe*. Montevideo: Imprenta 'El Siglo Ilustrado', 1907.

Azuma, Uichirô, "Note on Mr. E. Loedel-Palumbo's paper «Die Form der Raum-Zeit-Oberfläche eines Gravitationsfeldes»", *Tohoku Mathematical Journal*, 29 (Sendai, 1928): 158-59.

Besio Moreno, Nicolás, "Universidad Contemporánea. Evolución necesaria de la Universidad Argentina. Los estudios utilitarios, la investigación pura y la ciencia desinteresada". (Discurso de incorporación del 3 de octubre de 1923), *Anales de la Academia Nacional de Ciencias Exactas, Físicas y Naturales de Buenos Aires*, I (Buenos Aires, 1928): 101-117. [Originalmente publicado en Buenos Aires: *Anales de la Sociedad Científica Argentina*, 106: 101-117].

Broggi, Ugo, 'Die Axiome der Wahrscheinlichkeitsrechnung', Tesis doctoral, Univeristät Göttingen, 1907.

Broggi, Ugo, *Análisis Matemático*, La Plata: UNLP, Facultad de Ciencias Físicas, Matemáticas y Astronómicas, 1919-1927.

Bunge, Mario, *Causality: The Place of the Causal Principle in Modem Science*, Cambridge, Mass., Harvard University Press, 1959.

Colegio Libre de Estudios Superiores, *Quince años de Labor, 1930-1945*, Buenos Aires, Talleres Gráficos Emilio Bustos, 1946.

Collo, José B., Isnardi, Teófilo y Aguilar, Félix, "Teoría de la relatividad", *Boletín del Centro Naval*, 41 (Buenos Aires, 1923-1924): 263-284 (no. 442), 413-449 (no. 443) y 747-762 (no. 445).

Daniell, A., "Abstract No. 1498", *Physics Abstracts*, 31 (Londres, 1928): 435.

Deulofeu, Venancio; Galloni, Ernesto E. y Santaló, Luís A., "Historia de la Academia". *Anales de la Academia Nacional de Ciencias Exactas, Física y Naturales*, 27 (Buenos Aires, 1975): 149-273.

Eisenstaedt, Jean, "The early interpretation of the Schwarzschild solution". Don Howard y John Stachel (eds.), *Einstein and the history of General Relativity*, Boston, Birkhäuser, 1989: 213-233, (en p. 229).

*El Argentino*, "Desde ayer desempeña la Presidencia de la Universidad Nacional platense un ex-alumno", *El Argentino*, Año 22/8355, (La Plata, 2 diciembre, 1927): 2.

Estrada, Dardo, *Historia y bibliografía de la imprenta en Montevideo 1810-1865*. Montevideo, Cervantes, 1912.

Ewles, J., "Abstract", *Physics Abstracts*, 29 (Londres, 1926): 628.

Fermi, Enrico, *Conferencias*. Buenos Aires, Universidad de Buenos Aires, Facultad de Ciencias Exactas, Físicas y Naturales, 1934, Serie B, Publicación No. 15.

Gangui, Alejandro y Ortiz, Eduardo L., "Marzo-abril 1925: Crónica de un mes agitado: Albert Einstein visita la Argentina". *Todo es Historia*, 454 (Buenos Aires, 2005): 22-30.

Gangui, Alejandro y Ortiz, Eduardo L., "Einstein's Unpublished Opening lecture for his Course on Relativity Theory in Argentina, 1925". *Science in Context*, 21/3 (Cambridge, 2008): 435-450.
https://doi.org/10.1017/s0269889708001853

Gangui, Alejandro y Ortiz, Eduardo L., "First echoes of relativity in Argentine astronomy", G. Romero, S. Cellone and S. Cora (eds.), *Historia de la Astronomía* AAABS No. 2 (suplemento), La Plata, Argentina, 2009: 31-37.

Gangui, Alejandro y Ortiz, Eduardo L., "Anti-positivismo, ciencias teóricas y relatividad en la Argentina de la década de 1920". *Revista Brasileira de História da Ciência*, 4/2 (Rio de Janeiro, 2011): 201-18.

Gangui, Alejandro y Ortiz, Eduardo L., "The scientific impact of Einstein's visit to Argentina, in 1925". *Asian Journal of Physics*, 23/1 (Bangalaru, 2014): 81-90.

Gans, Richard, "Asymmetrie von Gasmolekeln. Ein Beitrag zur Bestimmung der molekularen Form", *Annalen der Physik*, 370/10 (Berlin, 1921): 97-123.

Gaviola, Enrique, "Homenaje a Ricardo Gans en su 70° aniversario: Introducción". *Revista de la Unión Matemática Argentina*, 14/3 (Buenos Aires, 1950): 100-108.

Hansen, Octavio, "Preparatorios para los estudios de Ingeniería", *Anales de la Universidad*, 17/21 (Montevideo,1912): 39-48.

Jiménez Landi, Antonio, *La Institución Libre de Enseñanza. Los orígenes*. Madrid, Taurus, 1973.

Kasner, Edward, "The impossibility of Einstein fields immersed in flat space of five dimensions". *American Journal of Mathematics*, 43 (Baltimore, 1921a): 126-129.

Kasner, Edward, "Finite representation of the solar gravitational field in flat space of six dimensions". *American Journal of Mathematics*, 43 (Baltimore, 1921b): 130-133.

Kasner, Edward, "Geometrical theorems on Einstein's cosmological equations", *American Journal of Mathematics*, 43 (Baltimore, 1921c): 217-221.

Kasner, Edward y Newman, James R., (1940). *Mathematics and the Imagination*, New York. Simon and Schuster, 1940.

Loedel, E., *Manual del sistema métrico de pesos y medidas, exposición completa, teórica y práctica*, Montevideo, El Autor, 1864.

Loedel Palumbo, Emilia Z., "Entrevista con la Ing. Emilia Z. Loedel Palumbo". (Montevideo, *Mundo Uruguayo*), 15 septiembre, 1927.

Loedel Palumbo, Enrique, *Nuevos conceptos y aplicaciones sobre algunos puntos de física*. Montevideo, Imprenta El Siglo Ilustrado, 1920.

Loedel Palumbo, Enrique, "Las constantes ópticas de la molécula de sacarosa. Su «forma geométrica»". *Contribución al estudio de las ciencias físicas y matemáticas,* 5 (La Plata, 1926a): 53-78.

Loedel Palumbo, Enrique, "Optische und elecktrische Konstanten des Rohrzuckers". *Annalen der Physik*, 384 (Berlin, 1926b): 533-549. [Con un apéndice de Gans].

Loedel Palumbo, Enrique, "Forma de la superficie espacio-tiempo, de dos dimensiones, de un campo gravitacional proveniente de una masa puntiforme". *Contribución al estudio de las ciencias físicas y matemáticas*, 4/73 (La Plata, 1926c): 79-87.

Loedel Palumbo, Enrique, "Die Form der Raum-Zeit-Oberfläche eines Gravitationsfeldes, das von einer punkt-förmigen Masse herrürt", *Physikalische Zeitschrift*, 27 (Berlin, 1926d): 645-648.

Loedel Palumbo, Enrique, "La velocidad de la luz en un campo gravitacional", *Contribución al estudio de las ciencias físicas y matemáticas*, 4 (La Plata, 1926e): 455-460.

Loedel Palumbo, Enrique, "Estructura del átomo". *Cursos y Conferencias*, Nos. 1, 2: Estructura del átomo, No. 4: Modelo Atómico de Bohr, No. 5: Espectros Roentgenianos y No. 8: Espectros Roentgenianos y estructura del átomo. (Buenos Aires, Editorial CLES, 1931).

Loedel Palumbo, Enrique, "Lógica y Metafísica (Una introducción al estudio del problema de la causalidad)". *Cursos y Conferencias*, 9, Nos. 131-132. (Buenos Aires, Editorial CLES, 1934).

Loyarte, Ramón G., "La nueva mecánica atómica", *Contribución al estudio de las ciencias físicas y matemáticas,* 11, 4 (La Plata, 1927): 137-167.

Möglich, F. K. S., "Die Form der Raum-Zeit-Oberfläche eines Gravitationsfeldes, das von einer punkt-förmigen Masse herrürt". *Jahrbuch der Mathematik,* (Berlin, 1926): 52.0931.02.

Mussio, Agustín H., "Sepelio del Ing. Octavio Hansen", *Renacimiento*, 3/72, (Montevideo, noviembre 1926).

Nicolai, Georg Friedrich. *Die Biologie des Krieges*, Zurich, Orell Füssli, 1917.

Nicolai, Georg Friedrich. *Biología de la guerra*, con un prólogo de Romain Rolland, (en traducción de Diego Abad de Santillán)**.** Buenos Aires, Editorial CLES, 1932.

Ortiz, Eduardo L. "A convergence of interests: Einstein's visit to Argentina in 1925". *Ibero-Americanisches Archiv*, 20 (Berlin, 1995): 67-126.

Ortiz, Eduardo L., "The transmission of science from Europe to Argentina and its impact on literature: from Lugones to Borges", E. Fishburn y E. L. Ortiz (eds.), *Borges and Europe revisited*, London, University of London, 1998: 108-123.

Ortiz, Eduardo L., "Julio Rey Pastor, su posición en la escuela matemática argentina", *Revista de la Unión Matemática Argentina*, 52/1 (Buenos Aires, 2011a): 149-194.

Ortiz, Eduardo L. "The emergence of Theoretical Physics in Argentina, Mathematics, Mathematical Physics and Theoretical Physics, 1900-1950". L. Brink and V. Mukhanov, (eds.), *Remembering Héctor Rubinstein*, Trieste: SISSA, 2011b: 13-34.

Ortiz, Eduardo L. "Maniobras Científicas- Ciencia, Investigación y Defensa", en G. A. Visca (editor), *La Ingeniera Militar*, Buenos Aires: Dunken, pp. 21-64, 2015.

Ortiz, Eduardo L., "Mario Bunge in the Complex Argentina of the 1940s-1960s", M. R. Matthews (ed.), *Mario Bunge: A Centenary Festschrift*, Capítulo 3, Berlin: Springer-Verlag, 2019.

Ortiz, Eduardo L., *Doing physics in the Pampas: The Argentine correspondence de Emil Bose (1909-1911)*, London, The Humboldt Press, in press 2021.

Ortiz, Eduardo L. y Otero, Mario H., "Removiendo el ambiente: La visita de Einstein al Uruguay en 1925". *Mathesis,* 2/1 (Mexico, 2001). 1-35.

Premier Congrès International de Philosophie Scientifique. *Premier Congrès International de Philosophie Scientifique*, Actualités Scientifiques, fascicules 388-395, Paris, Hermann, 1936.

Pyenson, Lewis y Singh, M., "Physics on the periphery; a world survey, 1920-1929", *Scientometrics*, 6(5) (Berlin, 1984): 279-306. https://doi.org/10.1007/bf02020129

Ramón y Cajal, Santiago, *Reglas y consejos sobre investigación científica*, Madrid, Fortanet, 1899.

Ricaldoni, Pedro, *Nociones elementales de Física Popular ajustadas a las clases elementales superiores y técnico-elementales*. Montevideo, Imprenta El Telégrafo Marítimo, 1874.

Ricaldoni, Pedro y de la Vega, Carlos, *Compendio del sistema métrico decimal acompañado de un Tratado completo de Aritmética práctica*, Montevideo, Imprenta y Tipografía a vapor, 1864.

Rinke, Stefan, "German migration to Argentina (1918-1933)", Thomas Adam, *Germany and the Americas*, Santa Bárbara, ABC-Clio, 2005, I: 27-31.

Sagastume Berra, Alberto E. y Grinfeld, Rafael, 'Mecánica atómica'. *Anales de la Sociedad Científica Argentina* (Buenos Aires, 1928-1930): 11-42 (no. 105), 7-24 y 159-177 (no. 106); 209-238 (no. 109).

Sagastume Berra, Alberto E. y Grinfeld, Rafael, *Mecánica atómica*, Buenos Aires: Coni/Universidad Nacional de La Plata, 1930.

Schmitz, François, *Le cercle de Vienne*, Paris, Vrin, 2009.

Social Progress, "Social Progress (Uruguay)". *Bulletin of the Pan-America Union*, 52/116) (Washington, 1928): 112.

Sociedad de Amigos de la Educación Popular. "Memoria de la Comisión Directiva, Escrutinio de 1885". *Anales del Ateneo del Uruguay*, 4/10, no. 55 (Montevideo, marzo, 1886).

Thomas, J. S. G., "Abstract No. 312", *Physics Abstracts*, 29, (Londres, 1926): 87-88.

Villacorta Baños, Francisco, *El Ateneo de Madrid, círculo de convivencia intelectual (1883-1913)*, Madrid: Consejo Superior de Investigaciones Científicas, 1978.

Westerkamp, José Federico, *Evolución de las Ciencias en la República Argentina, 1923-1972, La Física, Vol. II*. Buenos Aires, Sociedad Científica Argentina, 1975.

Xenus, "Un libro interesante". *La Razón*, (Montevideo, 24 de noviembre 1920).

Zum Felde, Alberto, *Proceso Intelectual del Uruguay*, Vols. I-III, Montevideo, Ediciones del Nuevo Mundo, 1987

# The physicist Enrique Loedel Palumbo in the Montevideo-Buenos Aires-La Plata scientific corridor: 1920-1930

*In this paper we consider Montevideo's liberal progressive atmosphere towards the end of the nineteenth century and, within it, the trajectory of young science student Enrique Loedel Palumbo. We discuss some of his activities in Argentina, where he moved to study Physics at a new and well-equipped Physics Institute at the National University of La Plata, where he would later become a leading figure. Initially Loedel Palumbo worked on the structure of complex molecules based on their magnetic, electric and optical properties. Later, when Einstein visited Argentina in 1925, he had chance to exchange ideas, which led to the publication of the first of a series of papers on the Theory of Relativity in some of Germany's leading scientific journals. Loedel Palumbo integrated fully with the intellectual life of his adopted country, becoming one of Argentina's top physicists and philosophers of science of his day. He is a valuable example of the deep intertwining of the intellectual life of Uruguay and Argentina in the first-half of the twentieth century. Like his Argentine*

*colleagues, he also experienced the consequences of living through a complex historical period.*